\documentstyle[12pt,equation]{article}
\begin{document}
\pagestyle{empty}
\def\to{\rightarrow}
\def\rs{\mbox{$\sqrt{s}$}}
\def\pt{\mbox{$p_T$}}
\def\ccbar{\mbox{$c \bar c$}}
\def\f2gam{\mbox{$F_2^\gamma$}}
\def\gstarga{\mbox{$\gamma^* \gamma$}}
\def\gamgam{\mbox{$\gamma \gamma$}}
\def\eplem{\mbox{$e^+e^-$}}
\def\glgam{\mbox{$g^\gamma (x,Q^2)$}}
\def\ep{\mbox{$ep$}}
\def\gamp{\mbox{$\gamma p$}}
\def\qgam{\mbox{$q^\gamma (x,Q^2)$}}
\def\fpigam{\mbox{$f_{P_{i}|\gamma}$}}
\def\fgame{\mbox{$f_{\gamma |e} $}}
\def\fqgam{\mbox{$f_{q|\gamma}$}}
\def\fggam{\mbox{$f_{g|\gamma}$}}
\def\pcsq{\mbox{$P_c^2$}}
\def\pc{\mbox{$P_c$}}
\def\fpie{\mbox{$f_{P_{i}|e} $}}
\def\fie{\mbox{$f_{i|e} $}}
\def\fqe{\mbox{$f_{q|e} $}}
\def\fge{\mbox{$f_{g|e} $}}
\def\fcgam{\mbox{$f_{c|\gamma}$}}
\def\fp2p{\mbox{$f_{P_2|p}$}}
\def\fgp{\mbox{$f_{g/p}$}}
\def\qsq{\mbox{$Q^2$}}
\def\psqmax{\mbox{$P^2_{\rm {max}}$}}
\def\psqmin{\mbox{$P^2_{\rm {min}}$}}
\def\psq{\mbox{$P^2$}}
\def\ftilde{\mbox{$\tilde f$}}
\def\frest{\mbox{$f^{\rm rest}$}}
\def\ben{\begin{subequations}}
\def\een{\end{subequations}}
\def\beq{\begin{eqalignno}}
\def\eeq{\end{eqalignno}}
\def\be{\begin{equation}}
\def\ee{\end{equation}}
\def\bea{\begin{eqnarray}}
\def\eea{\end{eqnarray}}
\renewcommand{\thefootnote}{\fnsymbol{footnote}}
\begin{flushright}
MADPH--95--895\\
June 1995\\
\end{flushright}
\hspace*{1cm}
\begin{center}
{\large\bf Multiple Interactions in $\gamma \gamma$ Collisions}\footnote{Talk
given at {\it Photon 95}, Sheffield, England, April 1995}\\
\vspace*{0.5cm}
Manuel Drees\footnote{Heisenberg Fellow}\\
{\it Univ. of Wisconsin, Dept. of Physics, 1150 University Avenue, Madison,
WI 53706}
\end{center}
\vspace*{0.9cm}
\begin{abstract}
The cross--section for two--photon events with (at least) two independent
partonic scatters is estimated, for LEP energies as well as a 500 GeV ``photon
collider". This results in events with (at least) four central (mini--)jets.
Such events might be found in existing data, and should be clearly seen at the
second stage of LEP.

\end{abstract}
\clearpage
\setcounter{page}{1}
\pagestyle{plain}
\normalsize\baselineskip=15pt
\setcounter{footnote}{0}

Most recent analyses of data on multi--hadron production in quasi--real
$\gamma \gamma$ collisions conclude \cite{1} that resolved photon interactions
\cite{2} contribute for partonic transverse momenta down to about 1.5 to 2.0
GeV. ALEPH finds \cite{3} $p_{T,{\rm min}} \simeq 2.5$ GeV for the DG
parametrization \cite{4}, but the same fit also gives a rather unusual form of
the soft $\gamma \gamma$ cross--section; the fitted value of $p_{T,{\rm min}}$
is strongly correlated to the size of the soft (``VDM") component. Further
support for $p_{T,{\rm min}}$ values in the vicinity of 1.5 GeV comes from
analyses of total cross--sections \cite{5} as well as event properties
\cite{6} in $p \bar{p}$ collisions.

The total {\em inclusive} cross--section for the production of (mini--)jets
grows quickly with the two--photon cms energy $W_{\gamma \gamma}$; for
$p_{T,{\rm min}}$ between 1.5 and 2.0 GeV, it will exceed the standard VDM
expectation for the total $\gamma \gamma$ cross--section for $W_{\gamma
\gamma} > 30$ to 100 GeV. The definition of an inclusive cross--section
contains a multiplicity factor; in this case the relevant multiplicity is that
of pairs of (mini--)jets. $\sigma^{\gamma\gamma}_{\rm jet} >
\sigma^{\gamma\gamma}_{\rm tot}$ then indicates that each $\gamma \gamma$
event contains on average more than one (mini--)jet pair. These additional jet
pairs are not produced by higher order QCD processes (gluon radiation);
rather, they are due to additional partonic interactions.

In the standard ``eikonalization" scheme \cite{5,6,7} one assumes partonic
scatters to occur independently of each other for fixed impact paramter $b$;
the total cross--section is then written as an integral over $b$:
\be
\sigma^{\gamma\gamma}_{\rm tot} = P^2_{\rm had} \int d^2b \left[ 1 -
e^{- A(b) \chi / P^2_{\rm had} } \right],
\ee
where the eikonal $\chi$ receives both non--perturbative and perturbative
contributions:
\be
\chi(W_{\gamma\gamma}) = \chi_{\rm soft}(W_{\gamma\gamma}) +
\sigma^{\gamma\gamma}_{\rm jet}(W_{\gamma\gamma}; p_{T,{\rm min}}).
\ee
A standard ansatz for the soft interactions is $\chi_{\rm soft} = \chi_0 +
\chi_1/W_{\gamma\gamma}$, where $\chi_{0,1}$ are constants. The parameter
$P_{\rm had}$ in eq.(1) describes \cite{7} the probability that a photon
undergoes transition into a hadronic state prior to the interaction. Finally,
the perturbative (mini--jet) contribution to the eikonal (2) not only depends
on the cut--off $p_{T,{\rm min}}$ but also on the parton densities in the
photon and on the QCD scale parameter; I will use the DG parametrization and
$\Lambda_{\rm QCD}=0.4$ GeV here, but other choices would do as well.

The cross--section for producing at least two pairs of (mini--)jets can be
computed by Taylor--expanding the exponential in eq.(1) \cite{8}:
\be
\sigma(\gamma \gamma \rightarrow \geq 2 \ {\rm jet \ pairs}) = \frac
{\left[ \sigma^{\gamma\gamma}_{\rm jet}(W_{\gamma\gamma}; p_{T,{\rm min}})
\right]^2} {\sigma_0},
\ee
with $1/\sigma_0 = \left[ \int d^2 b A^2(b) \right] / \left( 2 P^2_{\rm had}
\right)$. Note that the function $A(b)$ describing the transverse overlap of
the partons and the parameter $P_{\rm had}$ do not appear independently in
eqs.(1),(3); as explained in ref.\cite{8}, one can always compensate a change
of $P_{\rm had}$ by a simultaneous modification of the shape of $A(b)$. I
therefore fix $P_{\rm had} = 1/200$; I further assume that $A(b)$ is a
Gaussian centered at zero. The only free parameter is then the width of the
Gaussian or, alternatively, $\sigma_0$.

Fig.~1 shows predictions for the total hadronic $\gamma \gamma$ cross--section
using this simple ansatz, for $p_{T,{\rm min}}=1.6$ GeV and three different
values of $\sigma_0$. In order to keep the cross--section at $W_{\gamma\gamma}
=10$ GeV fixed, the constants $\chi_0$ and $\chi_1$ need to be changed along
with $\sigma_0$; I chose $\chi_0 = 375 \ (500,\ 1000)$ nb and $\chi_1 = 1.1 \
(2.5, \ 12.5) \ \mu$b$\cdot$GeV for $\sigma_0 = 650 \ (440, \ 275)$ nb. I do
not claim to make a significant prediction here, since I do not know how to
fix $\sigma_0$; rather, this figure is meant to demonstrate that the simple
ansatz used here can yield a reasonable total $\gamma \gamma$ cross--section
at high energies. In particular, for $\sigma_0 \simeq 300$ nb it more or less
follows the universal $s^{0.08}$ behaviour \cite{9}. I will therefore use 300
nb as default value of $\sigma_0$; eq.(3) shows that the signal for multiple
interactions simply scales like $1/\sigma_0$. Note finally that the inclusive
jet cross--section (dotted curve in Fig.~1) does indeed grow much more quickly
than a total hadronic cross--section is supposed to.

The evaluation of eq.(3) involves four integrals over parton densities in the
photon as well as two integrals over partonic transverse momenta; for
realistic applications one has two additional integrals over photon flux
factors. An explicit expression is given in 8. Note that eq.(3) as it
stands cannot be exactly correct, since energy--momentum conservation has not
been imposed. Unfortunately no unambiguous procedure to do this exists in the
framework of eikonalization. The simplest method, used in ref.\cite{8}, merely
imposes the constraint that the sum over all Bjorken$-x$ in one photon is less
than unity. Alternatively \cite{6} one can compute the cross--section for the
second interaction at the re--scaled $\gamma \gamma$ cms energy $W_{\gamma
\gamma}^{\rm resc.} = \sqrt{(1-x_1)(1-x_2)} W_{\gamma\gamma}$, where $x_{1,2}$
are the Bjorken$-x$ of the first hard interaction. The predictions of these
two modifications of eq.(3) are shown in Figs.~2 to 4 by the solid and long
dashed curves, respectively.

In these figures I have also required all jets to be reasonably central,
$|\eta_{\rm jet}|\leq 1.75$ at LEP (Figs.~2 and 3) and $\leq 2.0$ at a future
$\gamma\gamma$ collider based on a 500 GeV $e^+e^-$ linac with unpolarized
beams (Fig.~4). We see that these two predictions differ by some 20--30\% at
low energy and/or relatively large $p_{T,{\rm min}}$, but become very similar
once $p_{T,{\rm min}} \ll \sqrt{s}$. Again following ref.\cite{8},
$\sigma_{mn}$ in these figures stands for the cross--section for events with
(at least) $n$ (mini--)jets, $m$ of which pass the rapidity cut. Here only
results for $m=n$ are shown; as discussed in ref.\cite{8}, the combinations
$m=3,\ n=4$, and perhaps even $m=2, \ n=4$, might yield corrobating evidence
for multiple partonic interactions. For comparison, the two--jet inclusive
cross--section $\sigma_{22}$ is also shown. Note that only the twice resolved
contribution has been included here; inclusion of 1--res and direct
contributions would increase $\sigma_{22}$ very roughly by a factor of 2 at
LEP energies.

The conclusion of ref.\cite{8} was that the signal for events with multiple
partonic scatters might be marginal at TRISTAN. If the present estimate is to
be trusted, the situation should be much better already at LEP1, partly due to
the better angular coverage of LEP detectors, and partly due to the higher
beam energy. In particular, 100 pb$^{-1}$ of data should contain at least some
500 $\gamma \gamma$ events with at least four central partonic mini--jets with
$p_T > 2$ GeV. No jet isolation has been imposed here; neither have finite jet
reconstruction efficiencies been included. On the other hand, Figs.~2 and 3
also show that a substantial fraction of events containing at least four
central (partonic) jets actually contain at least a fifth central jet. This
can be deduced from the difference between the inclusive cross--section (solid
and long dashed), based on eq.(3), and the exclusive cross--section
(dot--dashed), where {\em exactly} two partonic scatters are required. This
latter cross--section is given by
\be
\sigma(\gamma \gamma \rightarrow 2 \ {\rm jet \ pairs} ) = \frac {1}
{2 P^2_{\rm had}} \left( \sigma^{\gamma\gamma}_{\rm jet} \right)^2
\int d^2 b A^2(b) e^{- A(b) \sigma^{\gamma\gamma}_{\rm jet} / P^2_{\rm had} }.
\ee
This differs from eq.(3) by the exponential factor, which gives the
probability that no additional hard interaction occurs after the first two
scatters; note that $\sigma^{\gamma\gamma}_{\rm jet}$ in the exponential has
again been computed at a rescaled (lower) value of $W_{\gamma\gamma}$. A
similar expression can be written for the exclusive 2--jet cross--section
(dotted curves in Figs.~2 to 4).

Notice that the probability for additional hard scatters, i.e. the ratio of
inclusive and exclusive cross--sections, is larger for events that already
contain two partonic interactions. The reason is that such events have
\cite{8} on average a considerably larger $W_{\gamma\gamma}$, as well as
smaller impact parameter $b$; recall that $\sigma^{\gamma\gamma}_{\rm jet}$
quickly increases with increasing $W_{\gamma\gamma}$, while $A(b)$ is peaked
at $b=0$. On the other hand, Figs.~3 and 4 show that the difference between
inclusive and exclusive cross--sections decreases very quickly with increasing
$p_{T,{\rm min}}$; even at the $\gamma\gamma$ collider this difference becomes
negligible for $p_{T,{\rm min}} > 5$ GeV. This is not surprsing, since
$\sigma_{44}$ itself drops just as quickly.\footnote{In the absence of
rapidity cuts, one has $\sigma_{22}({\rm incl.}) = \sigma_{22}({\rm excl.}) +
\sigma_{44}( {\rm incl.}) + \sigma_{44}({\rm excl.})$, up to terms that
involve at least four partonic scatters. This identity, which follows directly
from the definition of inclusive and exclusive cross--sections, is satisfied
to good approximation by my ansatz whenever the probability for additional
scatters is small, giving me some confidence that the scheme used here is at
least consistent. However, this identity does not hold in the presence of
rapidity cuts, which reduce $\sigma_{44}$ much more than $\sigma_{22}$.} On
the other hand, at such a collider most events with one hard interaction will
have a second partonic scatter with $p_T > 2$ GeV. This is demonstrated by the
lower dotted curve in Fig.~4, where such interactions have been forbidden; the
resulting cross--section is considerably smaller than the 2--jet inclusive
cross--section (short dashed curve).

Figs.~2 to 4 indicate that events with multiple (mini--)jets should be readily
observable at LEP1, and fairly common at LEP2; there the event rate should
remain observable even if all four jets are required to have (partonic) $p_T >
3$ GeV, which is a reasonably high value for two--photon physics. At a $\gamma
\gamma$ collider this $p_T$ cut could even be raised to 20--30 GeV. Of course,
an observable event rate will only yield a signal if the background from
higher order QCD $2 \rightarrow 4$ processes can be suppressed sufficiently. I
believe this should be possible at least for the lower range of $p_{T,{\rm
min}}$ values discussed here, but a quantitative study of this background is
clearly needed.

My estimates appear to yield significantly higher rates for events with
multiple interactions than found in the recent HERWIG study by Butterworth et
al. \cite{10}. This is probably largely due to the fact that they only allow
the non--perturbative contribution to the photonic parton densities (not to be
confused with the non--perturbative contribution to the eikonal $\chi$) to
eikonalize; this approach had earlier been used by Schuler and Sj\"ostrand
\cite{11}. On the other hand, the separation of the parton densities into
perturbative and non--perturbative contributions is not really well defined.
Moreover, in ref.\cite{12} it has been suggested that the perturbative
component should eikonalize as well, but with a narrower $A(b)$ distribution;
the argument here is that the system that develops from a hard $\gamma
\rightarrow q \bar{q}$ splitting is likely to be smaller than a
non--perturbative system like a (vector) meson. A narrower $A(b)$ increases
the probability for additional hard interactions. The simple one--component
ansatz (1) used here should fall somewhere in between these two extremes. Only
experiment can tell which (if any) of these ans\"atze is correct.

\vspace*{8mm}
\noindent
{\bf Acknowledgements}
I thank the authors of ref.\cite{10} for useful correspondence. This work was
supported in part by the U.S. Department of Energy under grant No.
DE-FG02-95ER40896, by the Wisconsin Research Committee with funds granted by
the Wisconsin Alumni Research Foundation, as well as by a grant from the
Deutsche Forschungsgemeinschaft under the Heisenberg program.

\clearpage

\end{document}